\documentclass[psamsfonts]{amsart}
\usepackage{latexsym,epsfig,amssymb,amsfonts,amsmath,graphicx,floatflt}
\def\opone{\leavevmode\hbox{\small1\kern-3.8pt\normalsize1}}

\newcommand{\ket}[1]{\left | \, #1 \right \rangle}

\newcommand{\braket}[2]{\left\langle\, #1\,|\,#2\,\right\rangle}

\newcommand{\be}{\begin{equation}}
\newcommand{\ee}{\end{equation}}
\newcommand{\bea}{\begin{eqnarray}}
\newcommand{\eea}{\end{eqnarray}}
    \long\def\symbolfootnote[#1]#2{\begingroup%
    \def\thefootnote{\fnsymbol{footnote}}\footnote[#1]{#2}\endgroup} 
\newfont{\Bb}{msbm10}
\graphicspath{{Pictures/}}

\begin{document}
\title{Complex and unpredictable Cardano}
\author{Artur Ekert}
\address{Mathematical Institute, University of Oxford, Oxford OX1 3LB, UK.} 
\address{Centre for Quantum Technologies, National University of Singapore, Singapore 117543.}
\begin{abstract}
This purely recreational paper is about one of the most colorful characters of the Italian Renaissance, Girolamo Cardano, and the discovery of two basic ingredients of quantum theory, probability and complex numbers. The paper is dedicated to Giuseppe Castagnoli on the occasion of his 65th birthday. Back in the early 1990s, Giuseppe instigated a series of meetings at Villa Gualino, in Torino, which brought together few scattered individuals interested in the physics of computation. By doing so he effectively created and consolidated a vibrant and friendly community of researchers devoted to quantum information science. Many thanks for that!
\end{abstract}
\maketitle
\section{Gambling scholar}

\begin{floatingfigure}[r]{4.1cm}
\includegraphics[width=4cm]{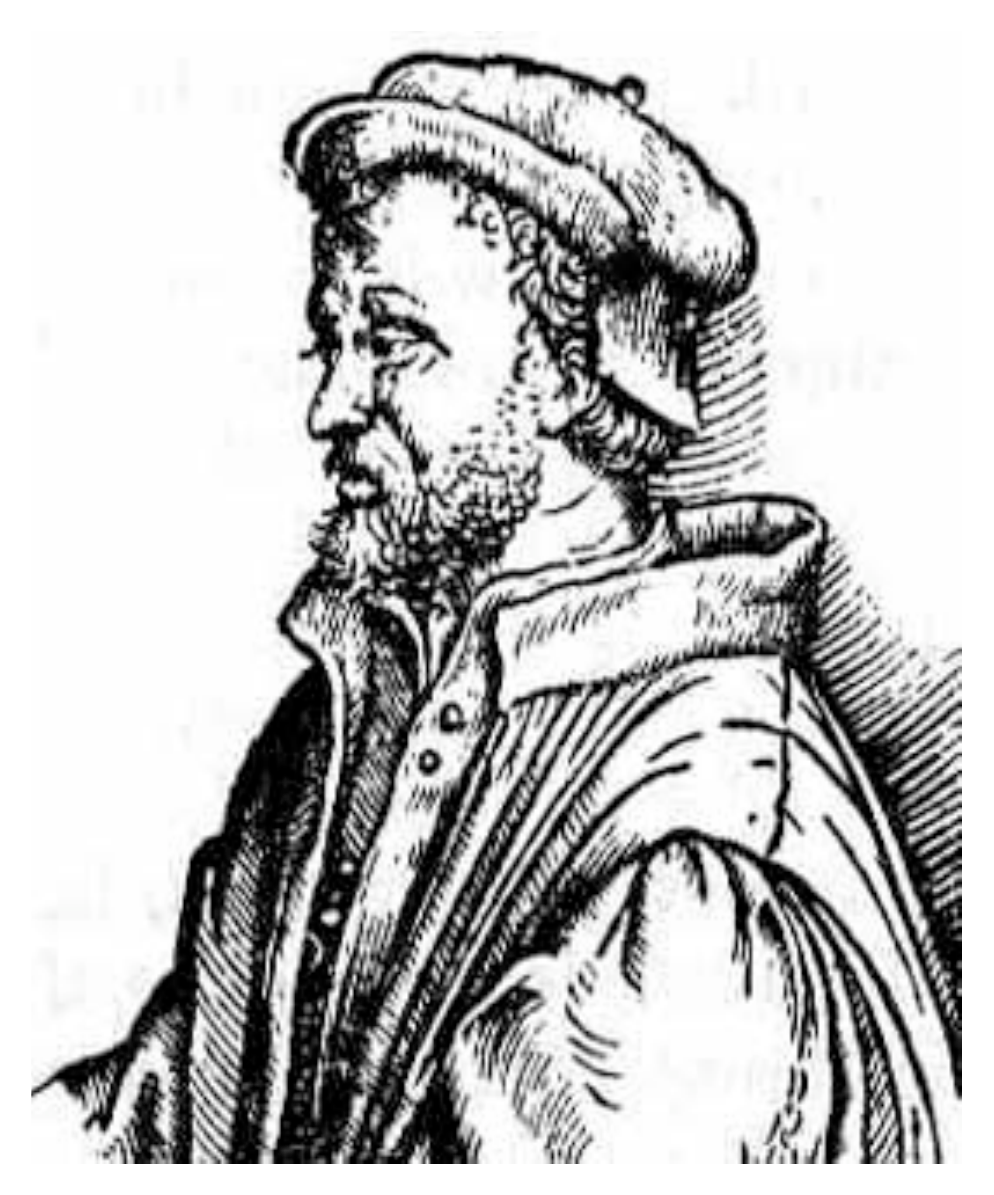}\\
\mbox{\qquad Girolamo Cardano}
\end{floatingfigure}
I always found it an interesting coincidence that the two basic ingredients of modern quantum theory, namely probability and complex numbers, were discovered by the same person, an extraordinary man of many talents, a gambling scholar by the name of Girolamo Cardano.\footnote{Hieronymus Cardanus in Latin, Jerome Cardan in English.} He was born in 1501 in Pavia, rose to become one of the finest minds of his time, and died in squalor and solitude in 1576 in Rome. In his autobiography, \emph{De vita propria liber} (The book of my life), written when he was seventy four, he described himself as ``...hot tempered, single minded, and given to woman,... cunning, crafty, sarcastic, diligent, impertinent, sad and treacherous, miserable, hateful, lascivious, obscene, lying, obsequious,.." and ``...fond of the prattle of old men."  In the chapter dedicated to ``stature and appearance" we learn that he was a man of medium height with narrow chest, long neck and exceedingly thin arms. His eyes were very small and half-closed, and his hair blond. He had high-pitched and piercing voice, and suffered from insomnia. He was afraid of heights and ``...places where there is any report of a mad dog having been seen". The narrative veers from his conduct, appearance, diet, and sex life to meetings with supernatural beings and academic intrigues. A patchy but surprisingly readable account of a mind-set of a Renaissance man.

Cardano loved writing and this being the Renaissance he wrote on just about everything, including medicine, astrology, philosophy, physics, mathematics and music. Many of his books were bestsellers read all over Europe.\footnote{It is quite remarkable how fast printing became profitable and books affordable. In about 1530, that is only 70 years after the invention of movable type, Cardano could purchase a printed pamphlet for the price of a loaf of bread and a copy of the New Testament for a daily wage.} When his extant works were collected and published, over eighty years after his death, they filled ten large folio volumes, which, even so, did not include everything he had written. His very popular semi-encyclopedic books, such as \emph{De subtilitate} and \emph{De rerum varietate}, published in the 1550s, covered a wide array of topics and included, apart from many anecdotes and speculations, descriptions of physical experiments and some ingenious mechanical devices. For example, he described a combination lock, a suspension consisting of rings inside one another that keeps an object in a fixed position regardless the motion of its support (gimbals), and a mechanical joint which facilitates transmission of rotation between two shafts positioned at various angles.
\begin{floatingfigure}[r]{4.2cm}
\includegraphics[width=4cm]{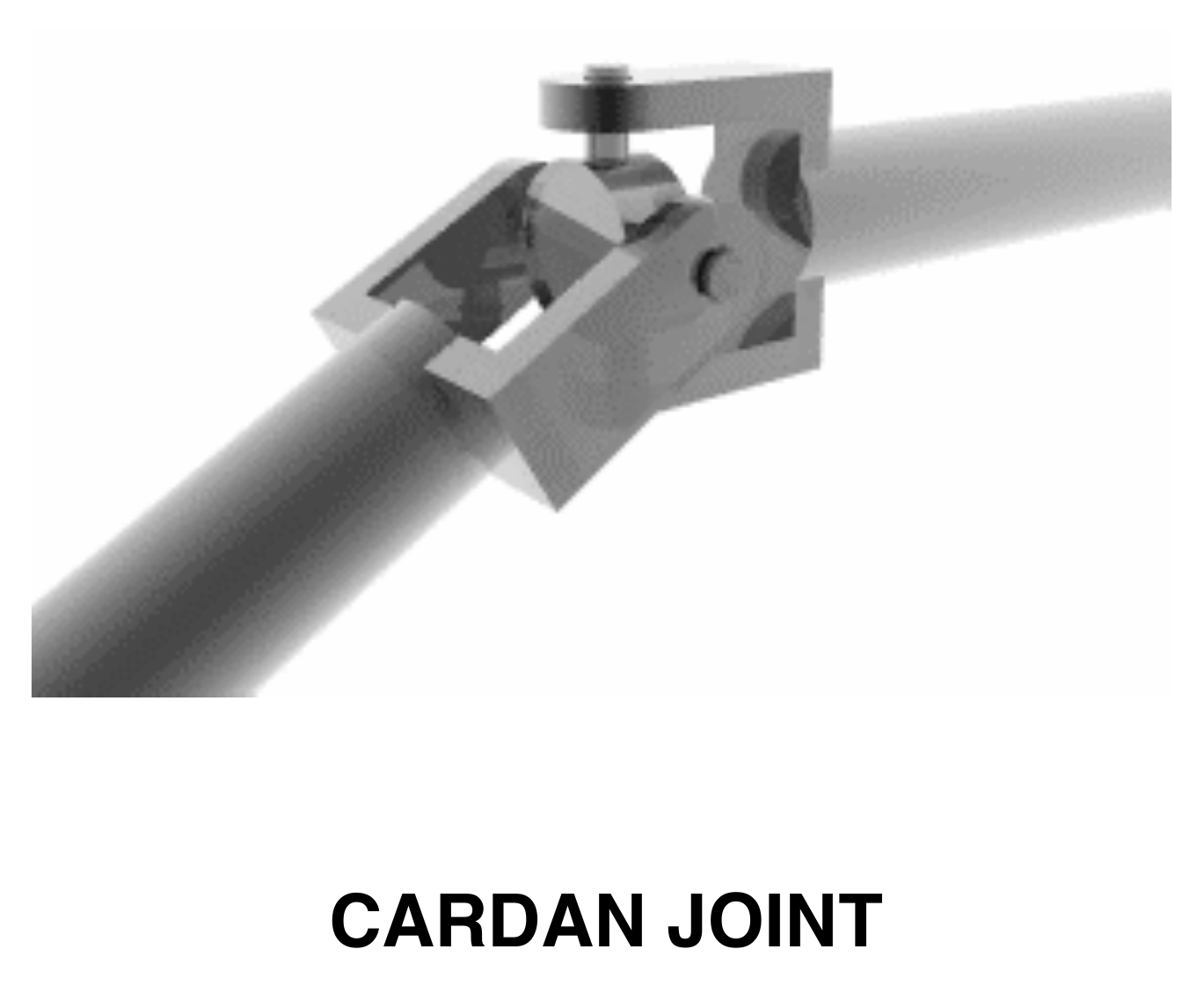}\\
%\mbox{\qquad {Cardan joint}}
\end{floatingfigure}
Cardano was the first to replace the pinhole in a camera obscura with a lens which resulted in brighter and sharper images. In cryptography he is known for his ``grille" --- a sheet of stiff paper with
few rectangular holes which was placed over the writing paper and a secret message was written in the holes. The grille was then removed, and an innocuous message was filled in around. In order to read the hidden message, an identical grille was placed over the text. Despite all these accomplishments, to his contemporaries Cardano was simply the most sought-after physician of his time, rivaled in fame only by his friend, the legendary anatomist Andreas Vesalius. When in 1552 Cardano undertook a long trip to St Andrews to treat the asthma of John Hamilton, the last Roman Catholic archbishop of Scotland,\footnote{In case you care ---  the cure was successful and the archbishop lived until 1571, when he was hanged, in full pontificals, for complicity of the murder of Lord Darnley, husband of Mary Queen of Scots} he was welcomed as a celebrity wherever he stopped. His company was sought by the good and the great of Europe, who were very keen on having their horoscopes drawn up by him.\footnote{In the 16th century Europe casting horoscopes was a common and widespread practice among skilled mathematicians and astronomers.} Even occasional flops, such as forecasting a long life for the English king, Edward VI (Henry VIII's son), who died one year after, did not tarnish his reputation. However, despite all these successes, his personal life was filled with tragedy, some of his own doing.

He was born to Fazio Cardano and Chiara Micheri. His father, a proud Milanese and a friend of Leonardo da Vinci, was a lawyer by training but his interests were mostly in mathematics and occult. His mother, Cardano wrote in his autobiography,  ``...was easily provoked...quick of memory and wit". An illegitimate, sickly and somewhat neglected child, he had to learn how to fend for himself. In 1520, encouraged by his father, he enrolled at the University of Pavia, about twenty miles south of Milan. Instructed in the three arts of the \emph{trivium}, that is grammar, logic and rhetoric, followed by the four mathematical arts of the \emph{quadrivium}, namely arithmetic, geometry, music and astronomy, he was on his way to study medical works of Hippocrates and Galen when yet another Italian war broke out. At the time Italy was in a total mess [and, according to Giuseppe, this has not changed since], effectively a playground for French and Spanish armies, with the rival Italian city states changing their allegiance according to their own parochial interests.\footnote{The spirit of the era was aptly captured by Orson Welles in ``The Third Man" (1949, directed by Carol Reed):  ``In Italy for thirty years under the Borgias they had warfare, terror, murder and bloodshed but they produced Michelangelo, Leonardo da Vinci and the Renaissance. In Switzerland, they had brotherly love, they had five hundred years of democracy and peace and what did that produce? The cuckoo clock." Well, as every German knows, the cuckoo clock was not invented in Switzerland, instead the Swiss produced the Bernoullis, Euler, Escher, Wilhelm Tell and Heidi.} Amid uncertainties and lack of funds the University of Pavia closed temporarily in 1523 and Cardano transferred to Padua, which was effectively the university of the Republic of Venice. He was already known for excesses at the gaming tables. Card games, dice and chess were the methods he used to make a living and he gradually started collecting his thoughts on chance and probabilities. Cardano knew that cheating at cards and dice was a risky endeavor, so he learned to win ``honestly" by applying his discoveries concerning probabilities. His \emph{Liber de ludo ale\ae} (The book on games of chance), is a compilation of his scattered writings on the subject, some of them written as early as 1525, some of them later, around 1565 or so. Unfortunately they were published posthumously in 1663, about nine years after a certain Chevalier de M{\'e}r{\'e}, an expert gambler, consulted Blaise Pascal (1623--1662) on some ``curious problems" in games of chance. Pascal wrote to his older colleague Pierre de Fermat (1601--1665), and it was through their correspondence, as we are often told,  the rules of probability were derived. That may be so, but let us do not forget that prior to Pascal and Fermat both Cardano and Galileo Galilei (1564--1642), and perhaps a few others, knew a thing or two about probability.

\section{Probability}

Of course, games of chance and the drawing of lots were discussed in a number of ancient texts and a number of mystics, loonies and mathematicians enumerated the ways various games can come out. The snag was, most of these enumerations were not enumerations of equally likely cases, so they could hardly be used to calculate odds in a systematic way. Cardano was more careful. In his writings he effectively defines probability as a number obtained by dividing the number of favorable cases by the number of possible cases. When discussing rolling a symmetric die Cardano states that there are six equally likely outcomes, and he goes on discussing rolling two, three or more dice. He correctly calculates the number of possible outcomes, i.e. $6\times 6$ for two dice, $6\times 6\times 6$ for three dice, etc., and the number of favorable outcomes for many problems of the type: three fair dice are rolled together, what is the chance that the total number of spots on the dice add up to nine or ten. This was related to an old gambling problem: I roll three dice and if the sum of spots is nine you win, if it is ten I win, and if it is neither I roll again. I offer you even odds. Is the game fair? I claim it is because there are six ways in which the sum can be nine
\be
1+2+6,\quad 1+3+5,\quad 1+4+4, \quad 2+2+5,\quad 2+3+4,\quad 3+3+3,
\ee
and there are also six ways for the sum to be ten,
\be
1+4+5,\quad 1+3+6,\quad 2+4+4, \quad 2+2+6,\quad 2+3+5,\quad 3+3+4.
\ee
At first glance the game looks fair, so we sit down and play, and after many rolls of dice you will notice that I am winning your money. Why? To see why you have to enumerate \emph{equally likely} outcomes. Imagine the three dice are clearly distinguishable and notice that the decomposition such as $1+3+5$ can occur in six different equally likely ways, namely
% \begin{center}
%{\dice 135}, {\dice 351}, {\dice 513},
 %\end{center}
 %\begin{center}
%{\dice 153}, {\dice 315}, {\dice 531}.
 %\end{center}
 \be
 (1,3,5),\; (3,5,1),\; (5,1,3,)\; (1,5,3),\; (3,1,5),\; (5,3,1).
 \ee
 
The decomposition such as $1+4+4$ can occur only in three equally likely
ways:
%\begin{center}
%{\dice 144}, {\dice 414}, {\dice 441}.
%\end{center}
 \be
(1,4,4),\; (4,1,4),\; (4,4,1).
 \ee
 
Finally the decomposition $3+3+3$ is realized by just one outcome:
%\begin{center}
%{\dice 333}.
%\end{center}
\be
(3,3,3).
 \ee

In general, when all the summands $a$, $b$ and $c$ are different we have six equally likely outcomes $(a,b,c)$, $(a,c,b)$, $(b,a,c)$, $(b,c,a)$, $(c,b,a)$, and $(c,a,b)$; when only two summands are different we have three equally likely outcomes   $(a,a,b)$, $(a,b,a)$, $(b,a,a)$, and when they are all the same we have just one outcome  $(a,a,a)$. We can now easily count the favorable outcomes in the game: $27$ outcomes give sum ten and only $25$ give sum nine. The probability the spots add up to nine is $25/216\approx 0.12$ and that they add up to ten is $27/216\approx 0.13$. The game is not fair and you should not accept the wager if I offer you even odds.\footnote{Could such a small difference have any impact on an actual game, when dice are hardly rolled more than one hundred times?}  Cardano formulated the principle that the stakes in an equitable wager should be in proportion to the number of ways in which each player can win, and he applied this principle to find fair odds for wagering with dice. In \emph{Liber de ludo ale\ae} he wrote

\begin{quote}
The most fundamental principle of all in gambling is simply equal conditions, e.g. of opponents, of bystanders, of money, of situation, of the dice box and of the die itself. To the extend to which you depart from that equity, if it is in your opponents favour, you are a fool, and if in your own, you are unjust.
\end{quote}
Cardano's definition of probability\footnote{Neither Cardano nor the French mathematicians of the 17th century used the word probability. I think the modern notion of probability, independent of gambling, was introduced by James Bernoulli in his \emph{Ars conjectandi}, published in 1713.} as the ratio of the number of favorable cases to the number of all possible cases is perfectly acceptable as long as you know (somehow) that all elementary outcomes are equiprobable. But how would you know? In many physical experiments the assumption of equiprobability can be supported by underlying symmetry or homogeneity. If we toss coins or roll dice we often assume they are symmetrical in shape and therefore unbiased. However, Cardano himself pointed out that ``every die, even if it is acceptable, has its favoured side". Today, casino dice are symmetrical to the precision of about $1/2000$ of a centimeter. Manufactures even take into account the tiny differences due to the different number of spots that are printed or embossed on each face. However, no matter how close a real object resembles a perfect Platonic die, for mathematicians this approach is far from satisfactory for it is circular - the concept of probability depends on the concept of equiprobability.

You may be surprised to learn that the search for a widely acceptable definition of probability took nearly three centuries and was marked by much controversy. In fact the meaning of randomness and probability is still debated today. Are there genuinely random, or stochastic, phenomena in nature or is randomness just a consequence of incomplete descriptions? What does it really mean to say that the probability of event $A$ is $0.75$?  Is this a relative frequency with which $A$ happens?  Or is it the degree to which we should believe $A$ will happen or has happened? Is probability objective or subjective?

Most physicists would \emph{probably} (and here I express my degree of belief) vote for objective probability. Indeed, physicists even \emph{define} probability as a relative frequency in a long series of independent repetitions. But how long is long enough? Suppose you toss a coin $1000$ times and wish to calculate the relative frequency of heads. Is $1000$ enough for convergence to ``probability" happen? The best you can say is that the relative frequency will be close to the probability of heads with at least such and such probability. Once again, a circular argument.  Richard von Mises (1883-1953), an Austro-Hungarian test pilot and a mathematician, tried to eliminate the problem by axiomatizing probability theory on the basis of frequencies in some \emph{special} infinite sequences of trials which he called ``collectives" but his arguments, in general, failed to convince sceptics. 

Karl Popper (1902-1994) tried to abandon the link with relative frequencies and make probability objective by viewing it as a sort of disposition, or propensity, that can be attributed to a single trial or experiment. For example, that a coin has probability of one half of showing up heads means that it has an internal tendency to show up heads in one half of the trials, and this property is attributed to the coin, even if the coin isn't actually tossed. However, what physically are those propensities? Popper was never able to give a convincing explanation.

In the subjectivist, sometimes called Bayesian\footnote{The term ``Bayesian" refers to a Presbyterian minister, Thomas Bayes (1702--1761), who proved a special case of what is now called Bayes's theorem. The theorem is often used to compute \emph{a posteriori} probabilities given new observations i.e.  to update existing beliefs in the light of new evidence.}, interpretation, probabilities reflect personal beliefs of ``rational" people.  Here rationality means conformity to a set of constraints on your preferences, e.g. if you prefer X to Y, and Y to Z, you must also prefer X to Z, and the like. When you say that your probability for a given event is $0.75$, I assume that you will back this up by betting on it and giving 3 to 1 odds. Moreover, I also assume that you are equally willing to take the other side of such a bet. One of the pioneers of this school of thought, Bruno de Finetti (1906--1985), refined his theory by betting with his students at La Sapienza on soccer games in the Italian League. He showed that a rational person must quantify his uncertainty in terms of probability or be susceptible to guaranteed losses in the betting game. 

Love it or loath it, subjectivism has proven to be influential especially among social scientists, Bayesian statisticians, and philosophers. However, if there are stochastic phenomena in nature, then they take place independently of whether anyone is looking at them and of how much anyone knows about them.  Thus they must be governed by some objective probabilities. But, as I have mentioned above, we can hardly specify what these objective probabilities are. There is a consensus though that once we know objective probabilities, propensities or whatever they might be, then we should, at the very least, set our subjective probabilities to be equal to the objective probabilities. Philosophers call this relationship between credence and chance the Principal Principle. Last but not least, quantum theory added an extra twist to the story. It had to. Probabilities are associated with measurements and what exactly distinguishes measurements from other physical processes [absolutely nothing]  has been much debated. I will comment on this later on. 

If you are prepared to forget about the meaning of probabilities and focus on the form rather than substance then the issue was resolved in the 1930's, when Andrey Nikolaevich Kolmogorov (1903--1987) put probability on an axiomatic basis in his monograph with the impressive German title \emph{Grundbegriffe der Wahrscheinlichkeitsrechnung} (Foundations of Probability Theory).

The mathematical approach introduces the concept of a sample space, which is a set  $\Omega$ of all possible elementary outcomes. For example, in the experiment of rolling a fair die once the elementary outcomes are $\Omega=\{1,2,3,... 6\}$. A more general outcome, or event, is defined as a subset of $\Omega$. The set of all subsets of $\Omega$ is called the power set of $\Omega$ and is denoted by $2^\Omega$. Elementary outcomes are special in that they cannot be decomposed into smaller subsets and any outcome can be composed out of elementary outcomes. According to Kolmogorov the probability function $\Pr$, also called a probability distribution or just probability, is any function on the subsets of $\Omega$,
\be
\Pr : 2^\Omega\mapsto [0,1]
\ee
which satisfies the following three conditions
\be
\Pr (\emptyset)=0, \qquad \Pr (\Omega)=1,\qquad \Pr (A\cup
B)=\Pr(A)+\Pr(B)\; \mbox{ if } \;A\cap B=\emptyset .
\ee

That's it. These are the Kolmogorov axioms that underpin the mathematical probability theory. Let us add a couple of definitions. If $A$ is possible, i.e. if $\Pr(A)\ge 0$, then we call
\be
\Pr (B|A)=\frac{\Pr(A\cap B)}{\Pr(A)}
\ee
the conditional probability of $B$ given $A$. We say that $A$ and $B$ are independent if $\Pr (B|A)=P(B)$, which implies $\Pr(A\cap B)=\Pr(A)\Pr(B)$. We can now study these axioms and their consequences for their own sake and this is what a pure mathematician does for a living. The axioms take care of mathematical consistency and tell you how to manipulate probabilities once you have the numbers. But how to get the numbers? Well, this brings us back to interpretations - from frequencies, from personal beliefs, from symmetries, you choose. No matter how you do it, once you identify all elementary outcomes and assign probabilities to them you may revert to mathematical formalism and calculate probabilities of more complicated outcomes using basically the additivity axiom. This deceptively simple rule for manipulating probabilities is so important for our future discussions that I state it again, but in a slightly modified form,
\begin{quote}
If any particular state of a physical system can be reached in several \emph{mutually exclusive} ways, $E_{1}$, $E_{2}$, $E_{3}$... , then the probability of the system ending up in this state is the sum of the probabilities of the constituent alternatives
\begin{equation}
p (E_1\cup E_2\cup E_3...)= p(E_1)+ p(E_2)+ p(E_3)... .
\end{equation}
\end{quote}

The additivity axiom makes lots of sense and agrees with our everyday intuition about chance and probability, however, as we shall soon see, many quantum phenomena do not conform to it. Indeed, there is no a priori reason why they should!

\section{Ars Magna}
We left Cardano in 1525 at a gaming table in Padua. At the time his reputation plummeted. He was regarded as a very rude and aggressive man. In 1526, after two rejections, he was awarded his doctorate in medicine. The first few years of his medical practice, in a small village near Padua, were difficult: he could hardly make a living, but eventually fortune smiled on him and, through some family connections, he found a teaching job in Milan. By then married, with children, he settled to a relatively comfortable life. Gradually he built up a medical practice and acquired influential patrons and patients. In 1536, his first medical book was published and in the very same year, a fifteen year-old Lodovico Ferrari entered his service as an errand boy. Cardano soon realized that he had acquired an exceptional servant. Lodovico was a mathematical prodigy and before long he became Cardano's most brilliant pupil.  True, his temper left a lot to be desired - once Ferrari came from a brawl with the fingers in his right hand cut off - but in years to come he would be Cardano's most loyal friend and disciple. The two became very close collaborators and went on to work out a general algebraic solution to the cubic and quartic equations, probably the most important mathematical achievement of the 16th century.

Today, with the benefit of our mathematical notation, we write the cubic equation as
\be
a x^3+b x^2+c x+d =0
\label{cubic}
\ee
with $a$, $b$, $c$ and $d$ being given real numbers, but not a single Renaissance mathematician would write it in this way.  At the time equations were usually described in words, for example, expression
\be
x^3+6x=20.
\ee
was written by Cardano as
\be
\mbox{cubus p.}\;  6.\; \mbox{rebus {\ae}qualis} \;20.\; ,
\ee
where the Latin~\emph{rebus} (``things") refers to unknown quantities. Moreover, negative numbers were not considered as proper numbers so different versions of this equation must have been written down depending on the signs of the coefficients.\footnote{Negative numbers were in use in ancient China, during the Han dynasty (206 BC--220 AD), but it was not until the 17th century that they were accepted in Europe. In Cardano's time, negative numbers were still treated with some suspicion, as it was difficult to understand what they really mean. This reminds me a story about a mathematician sitting in a cafe and watching an abandoned house across the street. After a while two people enter the house and a little later three people exit. How interesting - thinks the mathematicians - if a person enters the house it will be empty again.} No wonder it took a while to find a general solution of the cubic equation. Indeed, many tried and failed. A Franciscan friar, Luca Pacioli, in his influential book \emph{Summa de arithmetica, geometria, proportioni et proportionalita}, published in Venice in 1494, with beautiful illustrations by Leonardo da Vinci, maintained that such a solution could not be found. Then, sometime around 1515, Scipione del Ferro (1465--1526), a professor of mathematics at Bologna, found an algebraic formula for solving a specific cubic equation of the form $x^3+ cx= d$, with $c$ and $d$ positive. It was a real breakthrough, but del Ferro kept the solution secret.

It may sound strange in our publish-or-perish age, but keeping some mathematical discoveries secret was quite common at the time. Some mathematical tools were treated like trade secrets. After all, del Ferro and his colleagues made their living by offering their services to whoever paid them more. Patronage was hard to come by, there was no tenure, university positions were few and held by virtue of eminence and reputation, and challenges could come at any time. Scholars were often involved in animated public debates that attracted large crowds. These were very reminiscent of the knights tournaments in the Middle Ages with all the rituals - challenges, often publicly distributed in a form of printed pamphlets (\emph{cartelli}), responses (\emph{risposti}), witnesses, judges etc.  Town and gown, students and merchants, and all kind of spectators would gather in public squares or churches to watch the spectacle. The basic rule of combat was that no one should propose a problem that he himself could not solve. Reputations, jobs and salaries were at stake.

We do not know whether del Ferro ever used his result in a mathematical contest, but he was aware of its value, and shortly before his death in 1526, he passed the secret to his son-in-law, Annibale della Nave, and to one of his students, Antonio Maria Fiore. Although neither man published the solution, rumors began to spread that the cubic equations had been solved. In particular Niccol{\`o} Fontana of Brescia (1499 or 1500--1557), better known under his nickname Tartaglia (meaning the stammerer\footnote{His speech impediment was due to a saber cut to his mouth that he had received as a child from a French soldier. This was in 1512 when the French invaded Brescia - yet another military conflict during the messy period of the Italian Wars.}), boasted that he had discovered the solution to cubics of the form $x^3+b x^2=d$ (again $b$ and $d$ positive). At the time Tartaglia was a teacher of mathematics in Venice and Fiore, a native of Venice, was very keen on a good teaching job in his hometown. Confident in his mathematical abilities, Fiore challenged Tartaglia to a public contest. It was a bad idea. Tartaglia was a much better mathematician, and, as it happened, the night before the contest on the 13th of February 1535, he had figured out del Ferro's solution. The contest, held in Venice, was a humiliating defeat for Fiore and a great victory for Tartaglia. Overnight, an unknown teacher of mathematics from Venice became a nationwide celebrity. His star was rising. However, it ain't over 'til the fat lady sings. Enter Cardano.

When he heard of Tartaglia's triumph, Cardano asked him to reveal the secret and to give him permission to include the solution to cubic equations in~\emph{Practica arithmeticae generalis}, a book which he was preparing for publication. He promised to give full credit to Tartaglia, but Tartaglia categorically refused, stating that in due time he himself would write a book on the subject. When? He would not say, as he was preoccupied with his work on ballistics and translating Euclid's \emph{Elements} into Italian. Cardano did not give up. After several exchanges of letters Tartaglia accepted an invitation to visit Milan, possibly in the hope that through Cardano's connections he could secure a lucrative job with the Spanish governor in Milan.\footnote{Tartaglia was probably hoping that the governor might be interested in his pioneering work on artillery, ballistics, and designing fortifications.} The visit took place on the 25th of March 1539. This much we know for sure, but what exactly happened during the visit is not clear. We have two contradicting stories, one by Tartaglia and one by Ferrari. Tartaglia claimed he divulged the secret to Cardano in the form of an enigmatic  poem but only after Cardano had taken a solemn oath to keep the solution secret. Ferrari, who was present at the meeting, swore that Cardano took no oath of secrecy.  The word of one man stands against that of the other. 

One way or another, in May 1539 \emph{Practica arithmeticae generalis} appeared without Tartaglia's solution. However, Ferrari and Cardano were working hard and managed to extend Tartaglia's method to cover the most general case of the cubic equation. Ferrari went even further and worked out solutions to quartic equations. Meanwhile, Tartaglia still had not published anything on the cubics. In 1543, following rumors about the original discovery by del Ferro,  Cardano and Ferrari travelled to Bologna to meet Annibale della Nave. After examining dal Ferro's papers they found a clear evidence that twenty years earlier he indeed discovered the same solution as Tartaglia. Thus, even if Cardano had been sworn to secrecy the oath was no longer valid.

In 1545 Cardano's  \emph{Artis magn{\ae} sive de regulis algebraicis} (in short \emph{Ars Magna}) was published. It was a breakthrough in mathematics, a masterpiece comparable in its impact only to \emph{De revolutionibus orbium coelestium} by Copernicus and \emph{De humani corporis fabrica Libri septem} by Vesalius, both published two years earlier. In the book Cardano explores in detail the cubic and quartic equations and their solutions. He demonstrates for the first time that solutions can be negative, irrational, and in some cases may involve square roots of negative numbers. He acknowledges that he originally received the solution to the special cubic equation from Tartaglia. Chapter XI, titled \emph{De Cubo \& rebus {\ae}qualibus Numero} opens with the following line:  ``Scipio Ferro of Bologna well-nigh thirty years ago discovered this rule and handed it on to  Antonio Maria Fior of Venice, whose contest with Niccol{\`o} Tartaglia  of Brescia gave Niccol{\`o} occasion to discover it. He [Tartaglia] gave it to me in response to my entreaties, though withholding the demonstartion".\footnote{English translation by T. Richard Witmer from the 2007 Dover edition of \emph{The Rules of Algebra (Ars Magna)}.}  Cardano also credits Ferrari with the solution to quartic equations.
\begin{floatingfigure}[r]{5.2cm}
\includegraphics[width=5cm]{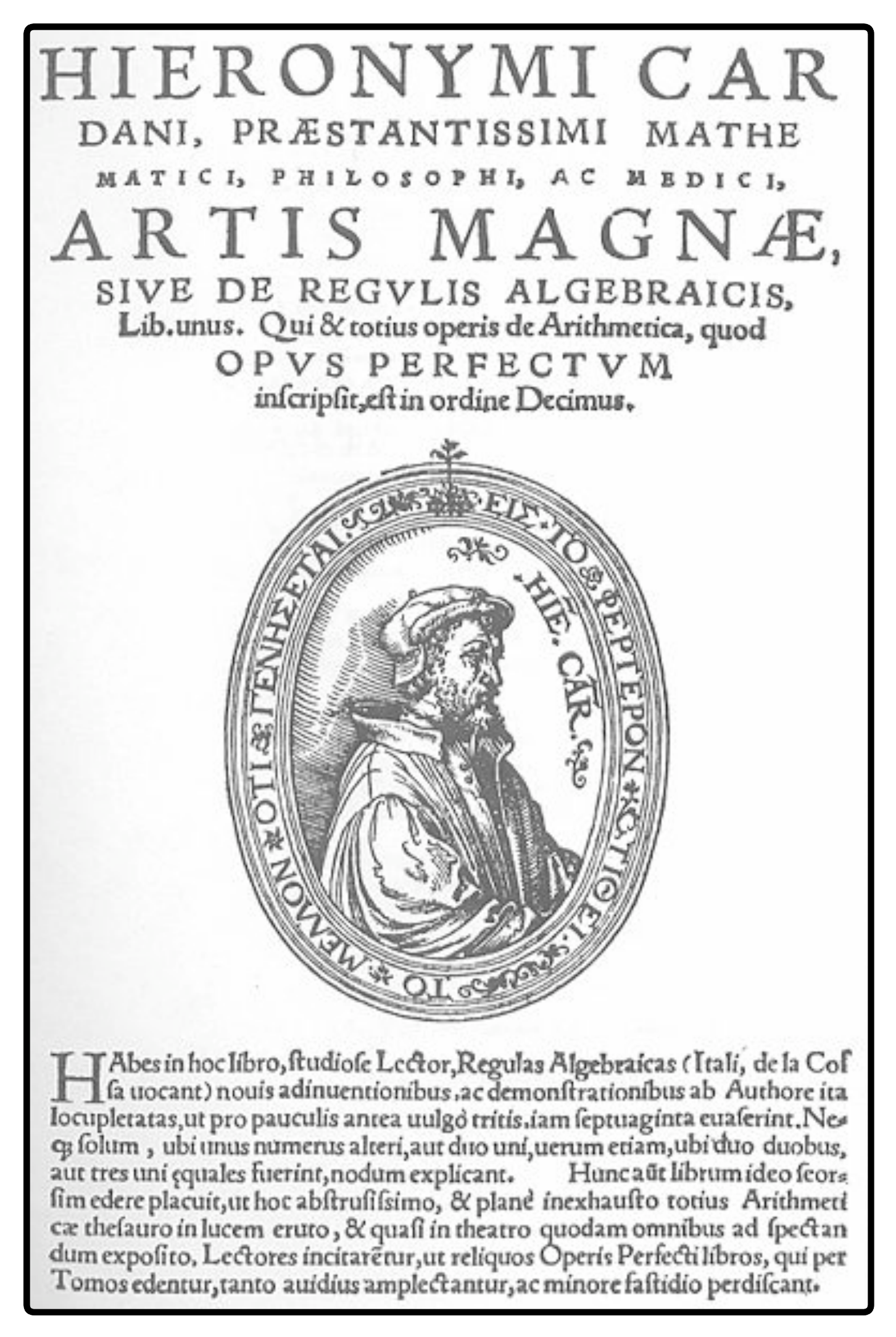}\\
\mbox{\;\;\;  Frontispiece of Ars Magna}
\end{floatingfigure}

After the publication of~\emph{Ars Magna} Tartaglia flew into a wild rage and started a campaign of public abuse directed at Cardano and Ferrari. He published his own work \emph{Quesiti et Inventioni Diverse} (New Problems and Inventions) which included correspondence with Cardano and what he maintained were word-by-word accounts of their meetings. They can hardly be regarded as objective, in fact, they read like a public rebuke. However, Cardano, who was now regarded as  the world's leading mathematician, couldn't care less. He did not pick up the fight, letting his loyal secretary, Ferrari, deal with it. And so Ferrari did. He wrote a \emph{cartello} to Tartaglia, with copies to fifty Italian mathematicians, challenging him to a public contest. Tartaglia however did not consider Ferrari as worthy of debate - he was after Cardano. Ferrari and Tartaglia wrote \emph{cartelli} and \emph{risposti}, trading insults for over a year until 1548 when Tartaglia received an offer of a good teaching job in his native town of Brescia. Most likely, in order to establish his credentials for the post, he was asked to take part in the debate with Ferrari. Tartaglia was an experienced debater and expected to win. The contest took place in the Church of Santa Maria del Giardino in Milan\footnote{It was a church of the Frati Zoccolanti (a branch of the Franciscans), built in 1455 at the current location of Piazza della Scala, rebuilt in 1582, and demolished in 1866.}, on the 10th of August 1548. The place was packed with curious Milanese, and the governor himself was presiding. Ferrari arrived with a large entourage of supporters, Tartaglia only with his own brother; Cardano was, conveniently, out of town. There are no accounts of the debate but we know that Tartaglia decided to flee Milan that night. Ferrari was declared an undisputed winner.

\section{Real roots, complex numbers}

Today, without any hesitation, we take the square root of $-1$ to get the imaginary unit denoted by $i$.\footnote{This notation was introduced in 1777 by the Swiss mathematician Leonhard Euler (1707--1783).} However, recall that back in the Renaissance even negative numbers were treated with a bit of suspicion, so taking roots of the suspicious numbers must have been almost heretical. Cardano did not mind. He came across the square roots of negative numbers while investigating the solution of the general cubic equation. He knew that the most general cubic equation~(\ref{cubic}) can be reduced to its ``depressed form",
\be
x^3 = px +q
\ee
by substitution $x\mapsto x- \frac{b}{3a}$. When working with this form it is helpful to visualize different situations which can occur by plotting the curves $x^3$ and $px+q$, and looking for the points of intersection. Depending on the signs and values of the coefficients $p$ and $q$ we find either one (diagram on the left) or three (diagram on the right) such points. They correspond to the cases in which the cubic equation has one or three real solutions.
\begin{center}
\includegraphics[width=12cm]{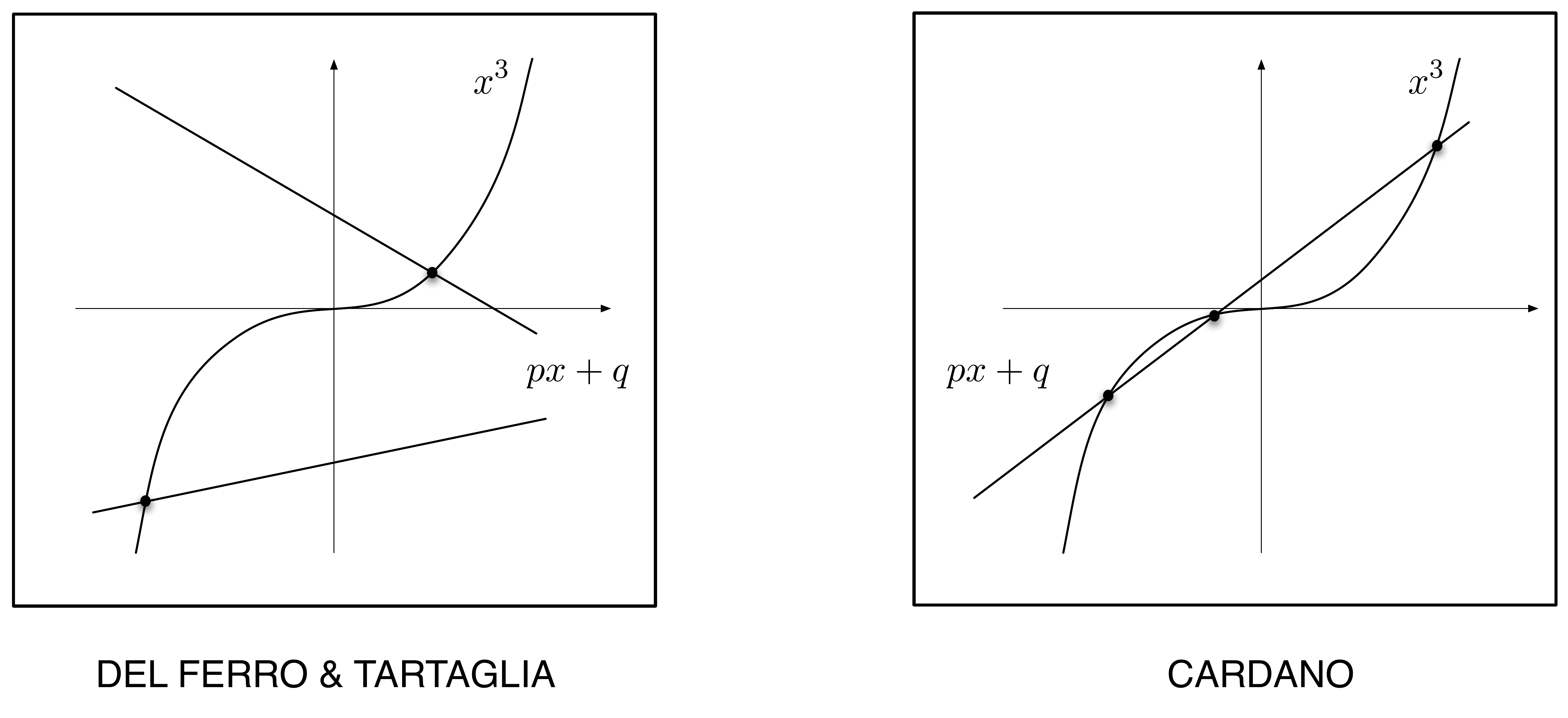}
\end{center}
Both del Ferro and Tartaglia knew how to find solutions in the cases corresponding to negative $p$ and possibly in some cases corresponding to positive $p$ and large absolute values of  $q$. These are the cases with one real solution. Cardano went further and figured out how to solve the so called \emph{casus irreducibilis} which occurs when
\be
\Delta=\left(\frac{q}{2}\right)^2 - \left(\frac{p}{3}\right)^3 <0
\ee
In this case the equation has three real solutions, as shown on the right diagram.  However, the general solution
\be
x=\sqrt[3]{\frac{q}{2}+\sqrt{\Delta}} + \sqrt[3]{\frac{q}{2}-\sqrt{\Delta}}
\ee
involves the square root of $\Delta$ ; hence $x$ is represented as the sum of two complex numbers, which are mutually conjugate, so the sum is always real. Cardano found the formula and realized that the solution to the cubic equation sometimes produced the square roots of negative numbers as an intermediate step. Take, for example, the equation
\be
x^3 = 15x+4
\ee
Cardano knew that $x=4$ was one of the solutions and yet it was a \emph{casus irreducibilis} as $\Delta=-121$ and the algebraic solution could only be expressed as
\be
x=\sqrt[3]{2+\sqrt{-121}} + \sqrt[3]{2-\sqrt{-121}}.
\ee
And how do you get $4$ out of that? Cardano was not happy about the appearance of square roots of negative numbers - he called them ``sophistic" - and remarked that it was a ``mental torture" to calculate with them. It was truly mysterious to him that starting with the real coefficients and ending with real solutions he had to traverse the new territory of imaginary numbers. In a later chapter of \emph{Ars Magna} he posed a problem of finding two numbers which sum to $10$ and such that their product is $40$. The solution is, of course, $5\pm\sqrt{-15}$, or rather
\begin{center}
5p: Rm: 15  and  5m: Rm: 15
\end{center}
which Cardano wrote down and commented that this result was as subtle as it was useless!

\section{Amplitudes}

Is there a connection between complex numbers and probabilities? Yes, there is. Amazingly enough they unite in the best physical theory we have today - a superb description of the inner working of the whole physical world - the quantum theory. 

Quantum theory asserts that probabilities are less fundamental than probability amplitudes, which are complex numbers $\alpha$ such that $|\alpha|^2$ are interpreted as probabilities. The Kolmogorov axioms of probability theory seem to codify our intuition about probabilities quite well, however, we have now an overwhelming experimental evidence that by manipulating probabilities alone we cannot describe our physical world. The main culprit is the additivity axiom - nature simply does not conform to it. Example? Here is a classic double-slit experiment that illustrates this.

Imagine a source of particles, say electrons, which are fired in the direction of a screen in which there are two small holes. Beyond the screen is a wall with a detector placed on it. If the lower hole is closed the electrons can arrive at the detector only through the upper hole. Of course, not all electrons will reach the detector, many of them will end up somewhere else on the wall, but given a location of the detector there is a probability $p_1$ that an electron emitted by the source reaches the detector through the upper hole. If we close the upper hole then there is a probability $p_2$ that an electron emitted by the source reaches the detector through the lower hole. If both holes are open it makes perfect sense to assume that each electron reaching the detector must have travelled either through the upper or the lower hole. The two events are mutually exclusive thus the total probability should be the sum $p=p_1+p_2$. However, it is well established experimentally that this is not the case.

\begin{center}
\includegraphics[width=10cm]{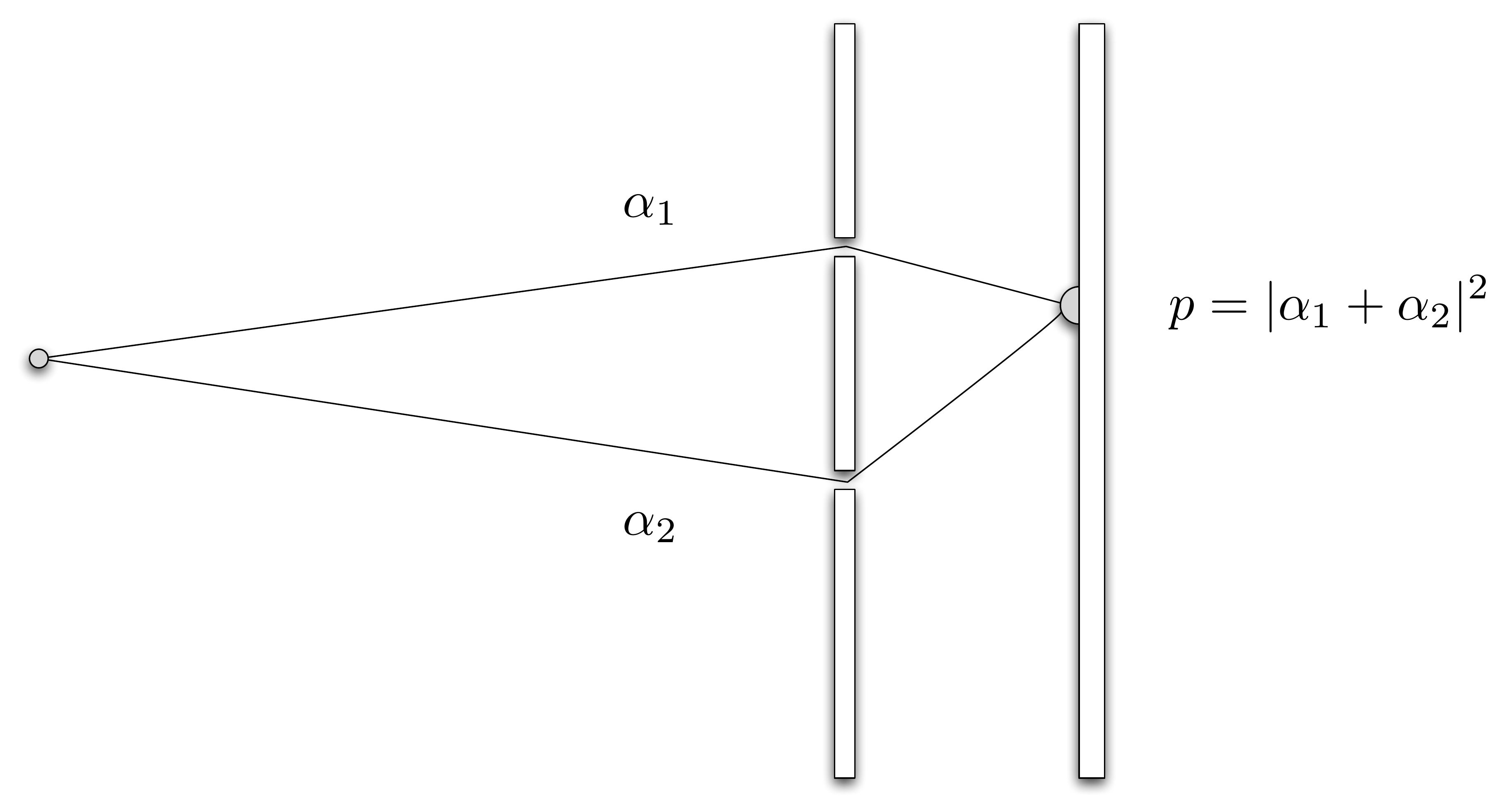}
\end{center}

Quantum theory asserts that the probability of an event is given by the square of the modulus of a complex number $\alpha$ called the probability amplitude. Thus we associate amplitudes $\alpha_1$ and $\alpha_2$ with the two alternative events , namely ``electron emitted by the source reaches the detector through the upper hole" and ``electron emitted by the source reaches the detector through the lower hole", respectively. For consistency we must require that  $|\alpha_1|^2=p_1$ and $|\alpha_2|^2=p_2$. However, and this makes quantum theory different, when an event can occur in several alternative ways, the amplitude for the event is the sum of the amplitudes for each way considered separately. In our case the amplitude that an electron reaches the detector when the two holes are open is
\be
\alpha = \alpha_1 + \alpha_2
\ee
and the associated probability
\bea
p=|\alpha|^2 =  |\alpha_1 + \alpha_2|^2 &=& |\alpha_1|^2 + |\alpha_2|^2
+ \alpha_1^\star\alpha_2 +  \alpha_1\alpha_2^\star\\
& = & p_1 + p_2 + |\alpha_1||\alpha_2|(e^{i(\theta_1-\theta_2)} +
e^{-i(\theta_1-\theta_2)})\nonumber \\
&=& p_1 + p_2 + 2 \sqrt{p_1 p_2} \; \cos (\theta_1-\theta_2).
\eea
where we have expressed the amplitudes in their polar form $\alpha_1=|\alpha_1|e^{i\theta_1}$ and $\alpha_2=|\alpha_2|e^{i\theta_2}$. The last term on the r.h.s.~marks the departure from the classical theory of probability. The probability of any two mutually exclusive events is the sum of the probabilities of the individual events, $p_1 + p_2$, modified by what is called the interference term, $2 \sqrt{p_1p_2}\;\cos(\phi_1-\phi_2)$. Depending on the relative phase $\phi_1-\phi_2$, the interference term can be either negative (destructive interference) or positive (constructive interference), leading to either suppression or enhancement of the total probability $p$.

Note that the important quantity here is the relative phase $\phi_1-\phi_2$ rather than the absolute values $\phi_1$ and $\phi_2$. This observation is not trivial at all. In simplistic terms - if an electron reacts only to the difference of the two phases, each pertaining to a separate path, then it must have, somehow, experienced the two paths. Thus we cannot say that the electron has travelled \emph{either} through the upper \emph{or} the lower hole, it has travelled through both. I know it sounds weird, but this is how it is.

Phases of probability amplitudes tend to be very fragile and may fluctuate rapidly due to spurious interactions with the environment. In this case, the interference term may average to zero and we recover the classical addition of probabilities. This phenomenon is known as \emph{decoherence}. It is very conspicuous in physical systems made out of many interacting components and is chiefly responsible for our classical description of the world -- without interference terms we may as well add probabilities instead of amplitudes.

Cardano had to go through the uncharted territory of complex numbers in order to obtain real solutions to cubic equations. As it happens, we do the same in quantum theory. We use complex amplitudes in order to calculate probabilities. The rules for combining amplitudes are deceptively simple. When two or more events are independent you multiply their respective probability amplitudes and when they are mutually exclusive you add them. This is just about everything you need to know if you want to do calculations and make predictions. The rest is just a set of convenient mathematical tools developed for the purpose of book-keeping of amplitudes. But, as auditors often remind us, book-keeping is important. Thus we tabulate amplitudes into state vectors and unitary matrices and place them in Hilbert spaces. We introduce tensor products, partial traces, density operators and completely positive maps, and often get lost in between.

\section{Amplitudes and probabilities revisited}

You may ask whether we have to use Cardano's discoveries. Can we describe the world without probabilities and complex numbers? Well, let us try. Suppose you want to construct a framework theory, a meta-level description of the world, by armchair reasoning alone. Just pour yourself a glass of good wine, take a seat, take a sip, and think. How would you like to have your theory? It will be raw, for sure. And it should be as simple as possible. To start, assume that there exist physical systems which evolve from one state to another. What is this evolution? Should it be stochastic? 

The concept of probability is useful no matter whether there are stochastic phenomena in nature or not. In the classical world, randomness arises as a consequence of incomplete description or knowledge of otherwise deterministic dynamics. Mind you, probability theory was developed by people who, by and large, believed that ``things don't just happen". Cardano, with all his superstitions, was the border-line case but two hundred years later Pierre-Simon Laplace (1749--1827) firmly believed that the world is ruled by causal determinism, i.e. every event is caused by, and hence determined by, previous events. Moreover, if at one time, we knew the positions and speeds of all the particles in the universe, then, at least in principle, we could calculate their behavior at any other time, in the past or future. This world view, known as predictive determinism, despite some practical difficulties, was basically the official dogma throughout the 19th century. It was challenged in the 20th century by quantum theory which ruled out sharp predictions of measurement outcomes. The predictive determinism is unachievable, no matter how much we know and how much computational power we have we cannot make precise predictions of what is going to happen. Thus we are stuck with probabilities. Your armchair theory better be a statistical theory.

Classical probability is a good starting point - let us see where it leads. Assume that any physical systems can be prepared in some finite number of distinguishable states. Introduce the state vector which tabulates probabilities of the system being in a particular state and make sure that admissible transformations preserve the normalization of probabilities. Given any vector $v$ with components $v_1, v_2,... v_n$ the $p$-norm of $v$ is defined as
\be
\left(|v_1|^p+|v_2|^p+... |v_n|^p\right)^{\frac{1}{p}}
\ee
thus for probability vectors you make sure the $1$-norm is preserved. Keep it simple, keep it linear, use transition matrices. Your admissible transformations are then represented by stochastic matrices $P$ - they have nonnegative elements such that $\sum_{m}P_{mn}=1$, i.e. entries in each column add up to one. The matrix element $P_{mn}$ is the probability that the system initially in state labeled by $n$ evolves over a prescribed period of time into the state labeled by $m$. The probability vector with components $p_n$ evolves as $p_n\mapsto \sum_n P_{mn}\; p_n$.

Take a sequence of two \emph{independent} evolutions, $P$ followed by $Q$. What is the probability that the system initially in state $n$ evolves over a prescribed period of time into the state $m$ via some intermediate state $k$? The evolutions are independent so for any particular $k$ the probability is $Q_{mk}P_{kn}$. But there are several intermediate perfectly distinguishable states $k$, thus there are several mutually exclusive ways to get from $n$ to $m$. Following the Kolmogorov additivity axiom you add up the constituent probabilities, $\sum_kQ_{mk}P_{kn}$, and discover that the matrix multiplication $QP$ in one swoop takes care of the multiplication and addition of probabilities. Products of stochastic matrices are stochastic matrices, so far so good. 

Now you add one more requirement - \emph{continuity} of evolution. Any product of stochastic matrices will give you a stochastic matrix but now you are asking for more. It should be possible to view any evolution as a sequence of \emph{independent} evolutions over shorter periods of time. In particular, we should be able to take the square root, or the cube root, or any root of any transition matrix and obtain a valid transition matrix. Take, for example, a physical system with two states, say a physical bit, and consider a transformation which swaps the two states; a logical {\tt NOT} if you wish, represented by the stochastic matrix,
\be
\begin{pmatrix}
0 & 1\\
1 & 0
\end{pmatrix}.
\ee
Take the square root. The two eigenvalues of this matrix are $\pm 1$ so you have to end with a matrix with complex entries, indeed

\be
\begin{pmatrix}
0 & 1\\
1 & 0
\end{pmatrix} = \frac{1}{ 2}\begin{pmatrix}
1+i & 1-i\\
1-i & 1+i
\end{pmatrix}\;
\frac{1}{2}\begin{pmatrix}
1+i & 1-i\\
1-i & 1+i
\end{pmatrix}
\ee

Square roots of stochastic matrices are, usually, not stochastic matrices. In other words - by adding the continuity requirement you gracefully thrashed your classical theory. Take another sip of wine and try again. 

Keep state vectors and transition matrices $T$ but let them have complex entries, simply because they pop up as soon as you start taking roots. Hopefully you will be able to relate complex numbers to probabilities later on. Take the continuity requirement seriously and parametrize transition matrices $T(t)$ with some real parameter $t$, that you may as well call time. Require that  $T(t+s)=T(t) T(s)$, for any two time intervals $t$ and $s$, and set $T(0)=\opone$. Ha! This clearly points towards an exponential map $T(t)=\exp(tX)$, where $X$ is any complex matrix. Now, taking the $n$th roots or inverses is a breeze: $T(t)^{1/n}=T(t/n)$ and  $T(t)^{-1}=T(-t)$. You also recall that any matrix can be written in its polar form $T=RU$ where $R$ is a positive matrix and $U$ is unitary, it is analogous to writing a complex number in the polar form or viewing linear transformation  $T$ as the ``stretching" $R$ and the ``rotation" $U$. But the exponential increase of stretching with time does not look good, you do not want to have exponential divergencies in your theory, so you had better drop $R$. Now, you are left with a unitary evolution of the form $U(t)=\exp{itX}$, where $X$ is Hermitian. It looks good, decent periodic evolution, no exponential divergencies. But what does it mean? Now, you have to follow your hunch - probability should be preserved under the admissible evolution, so what is it that remains invariant under unitary operations... Eureka! The length of a vector! The Euclidian norm or the $2$-norm, if you wish. Hence the squares of absolute values of complex components are probabilities. Now you have it all - state vectors with complex components, unitary transition matrices, and you know how to get probabilities out of the complex numbers. Congratulations, you guessed quantum theory without moving your butt from the armchair. Well, almost, there are a few holes in this plausibility argument, but they can be fixed, (with some more wine, of course). 

More refined arguments can be found in a number of papers, in particular, in a very readable exposition by Lucien Hardy~\cite{har01}, who argues, very convincingly, that if we try to construct a good statistical theory from a few (actually five) very reasonable axioms, then once we request \emph{continuity} of admissible evolutions  we will end up with quantum theory, and if this requirement is dropped we obtain classical probability theory. 

The connection between amplitudes and probability is not trivial. Even the pioneer, Max Born, did not get it quite right on his first approach. In the original paper proposing the probability interpretation of the state vector (wavefunction) he wrote~\cite{bor26}:  
\begin{quote}
...If one translates this result into terms of particles only one interpretation is possible. $\Theta_{\eta,\tau,m}(\alpha,\beta,\gamma)$ [the wavefunction for the particular problem he is considering] gives the probability$^*$ for the electron arriving from the $z$ direction to be thrown out into the direction designated by the angles
$\alpha,\beta,\gamma$... .

$^*$ Addition in proof: More careful considerations show that the probability is proportional to the square of the quantity $ \Theta_{\eta,\tau,m}(\alpha,\beta,\gamma)$.
\end{quote}

However, interestingly enough, the absolute values of amplitudes can also be interpreted as probabilities~\cite{bprob}. Why do we square the amplitudes? Born's rule does not have to be postulated, it follows from the formalism of quantum theory. Here we usually refer to Gleason's theorem~\cite{gle57}. Although very helpful in clarifying the formalism and telling us what follows from what, the theorem itself offers very little in terms of physical insights and has no bearing on the issue of what probability is. There are more interesting and more productive approaches. For example, Scott Aaronson added a nice computer science flavor to the whole story~\cite{aar04} by looking at a ``what if " scenario.  Suppose probabilities are given by the absolute values of amplitudes raised to power $p$. He showed that any linear operation that preserves the $p$-norm of a state vector is trivial apart from the two cases, namely, $p=1$ and $p=2$. For $p=1$ we get stochastic matrices, that is classical stochastic evolution, and for $p=2$ we get unitary matrices, that is quantum mechanics. In all other cases the only admissible operations are permutations of the basis vectors and sign changes and this may be not enough to account for our complex world.

As a realist - a true believer that science describes objective reality rather than our perceptions - I find Gleason's theorem too instrumental to my taste. After all, measurements are not just projectors but interactions between 
systems and measuring devices. Can unitary evolution alone shed some light on probability? Only then I would say that Born's rule really follows from the formalism of quantum theory. Fortunately, we do have a pretty good explanation of what probability really is in quantum physics. It was provided by David Deutsch~\cite{deu99}, with subsequent revisions by David Wallace~\cite{wal03}.\footnote{I would choose this approach, wouldn't I - one of  the authors was effectively my Ph.D. supervisor and another one my Ph.D. student, but I do not take any credit for this work.} They showed that no probabilistic axiom is required in quantum theory and that any decision maker who believes only in the non-probabilistic part of the theory, and is ``rational" in the sense we already described, will make all decisions that depend on predicting the outcomes of measurements as if those outcomes were determined by stochastic processes, with probabilities given by Born's rule. All this follows from the bare unitary evolution supplemented by the non-probabilistic part of decision theory! One may argue about the status of decision theory in physics, however, by any account this is quite a remarkable result. It shows that it does make sense to talk about probabilities within the Everett interpretation and that they can be derived rather than postulated. And it is beautiful to see how a deterministic evolution of a state vector generates randomness at the level of an observer embedded and participating in the evolution. 

Whatever the formalism, whatever the explanations, probability and complex numbers can hardly be avoided. It seems that we really need these Cardano's  discoveries.

\section{Epilogue}

Let us get back to 1548. What happened after the contest in the Church of Santa Maria del Giardino in Milan?

Tartaglia's appointment in Brescia was not renewed. He taught  there for about a year but his stipend was not paid. After many lawsuits, he returned, seriously out of pocket, to his previous job in Venice. He spent much of the rest of his life plotting and collecting a dossier against Cardano. He died in poverty in his house in the Calle del Sturion in Venice on the 13th of December 1557.

Ferrari became famous. He was appointed a tax assessor to the governor of Milan, and soon after retired as a young and rich man. He moved back to his hometown - Bologna - where he lived with his widowed sister Maddalena. In 1565 he was offered a professorship in mathematics at the university but, unfortunately, the same year he died of arsenic poisoning, most likely administered by his sister. Maddalena did not grieve much at his funeral and having inherited his fortune, remarried two weeks later. Her new husband promptly left her, taking with him all her dowry. She died in poverty.

Cardano outlived both Tartaglia and Ferrari. In 1546, fifteen years into his marriage, his wife died leaving Cardano the sole caretaker of his three children. This did not seem to affect him that much. He remained in Milan, lecturing in geometry at the University of Pavia, making money both as a physician and as a writer. From the day he published \emph{Ars Magna} till about 1560 he had everything he could possibly wish for - fame, position, money, respect. These were his golden years, but then, as it often  happens,  came difficult times. His eldest and most beloved son, Giambatista, married a woman who, by all accounts, was a despicable character. Publicly mocked and taunted about the paternity of his children he reached his limits of sanity and poisoned her. He was arrested. Cardano did everything he could for his son. He hired the best lawyers and paid all the expenses. Five doctors were brought in who stated that Giambatista's wife had not been poisoned, or at least, had not received a fatal dose. It seemed that a not-guilty verdict would be announced when, for some inexplicable reason, Giambatista confessed. He was sentenced to death, tortured, and beheaded in 1560. It was a real turning point in Cardano's life; he never recovered from his grief. He gave up his lucrative medical practice in Milan, took his baby grandson, Facio, with him, and moved to Pavia, where he became pathologically obsessed about his own safety. He reported a number of intrigues, attempts on his life and malicious accusations of professional incompetence and sexual perversion. In 1562, forced to resign his position in Pavia, Cardano, with some help from the influential Borromeo family, secured transfer to a professorship of medicine in Bologna. 

For a while Cardano's life resumed its old order; he was happy to touch base with his old friend Ferrari. However, problems with his children continued. Cardano's daughter, Chiara, died of syphilis, contracted as a result of her prostitution. His second son, Aldo, a perpetual thief, moved with him to Bologna but did nothing but drink and gamble. On a number of occasions frustrated Cardano had to bail him out of the staggering debts. Finally, when in 1569 Aldo gambled away all of his personal possessions and was caught stealing a large amount of cash and jewelry from his father, Cardano had him banished from Bologna. In 1570, Cardano himself was imprisoned for a few months by the Inquisition in Bologna. The charges against Cardano are not known. By the time one of his most vicious enemies, Tartaglia, was already dead and it is very unlikely he had anything to do with it. Rumor had it that Aldo was one of Cardano's accusers and, as a reward, he was appointed a public torturer and executioner for the Inquisition in Bologna.

On his release from prison, Cardano was sworn to secrecy about the whole proceedings and forbidden to hold a university post. He moved to Rome and appealed to the Pope\footnote{Gregory XIII - the one who ordered to reform the calendar}, who granted him a small pension and allowed him a limited practice in medicine. It was in Rome he started writing his autobiography \emph{De vita propia liber}. He died on the 21st of September 1576. Some say that Cardano predicted the exact date of his own death, some say that he starved himself to death to make this prediction true, we will never know.

 \section*{Sources}
While I would not want to guarantee that all the statements about Cardano that appear in this article are true (this is how you tell a scientist from a historian) I made an effort to consult the best sources I had access to. The following books were particularly useful to me and are recommended to anyone interested in Cardano's life. 
\begin{itemize}
\item Girolamo Cardano, \emph{The Book of My Life (De vita propia liber)}. Translated by Jean Stoner. New York Review Books, 2002.
\item Girolamo Cardano, \emph{The Rules of Algebra, (Ars Magna)}. Translated and edited by T. Richard Witmer. Dover Books on Mathematics, 2007.
\item Girolamo Cardano, \emph{Opera Omnia Hieronymi Cardani, Mediolanesis}, 10 vols, Spoon, Lyons, 1663. All 10 volumes can be downloaded from the website of the Philosophy Department, University of Milan (http://www. filosofia.unimi.it/cardano/). 
\item  {\O}ystein Ore, \emph{Cardano, the Gambling Scholar}. Princeton, 1953. Contains an English translation of \emph{Liber de ludo ale\ae} by Sydney H. Gould.
\item Henry Morley,  \emph{The life of Girolamo Cardano, of Milan, Physician} 2 vols. Chapman \& Hall, London 1854.
\item Arnaldo Masotti, \emph{Lodovico Ferrari e Niccol{\`o} Tartaglia. Cartelli di Sfida Matematica}, Ateneo di Brescia, Brescia 1974.
\item Florence N. David, \emph{Games, Gods and Gambling. A History of Probability and Statistical Ideas}. Dover Publications, 1998.
\item Nancy G. Sirasi,  \emph{The Clock and the Mirror: Girolamo Cardano and Renaissance Medicine.}  Princeton University Press,1997.
\item Anthony Grafton, \emph{Cardano's Cosmos: The Worlds and Works of a Renaissance Astrologer.} Harvard University Press, 2001.
\end{itemize}

 \section*{Acknowledgements}
My interest in Cardano goes back to my student days in Oxford and at the time I benefited greatly from a number of discussions with my colleagues mathematicians and historians, in particular with Roger Penrose, who was equally intrigued by Cardano's colorful personality and his role in discovering both probability and complex numbers. Discussions with David Deutsch, which rarely started before midnights, were always very useful in putting things in perspective. I am grateful to Charles Clark, L.C. Kwek and Rossella Lupacchini for their comments, corrections and help with the Italian and Latin sources. However, this paper is dedicated to my Italian friend Giuseppe Castagnoli with many thanks for our long walks and talks in Pieve Ligure, and never ending discussions over Piedmontese wine. My special thanks to Ferdinanda, Giuseppe's wife, for putting up with the two of us and preparing the best pesto in the known universe. 

%\bibliographystyle{plain}
%\bibliography{cardano}

\begin{thebibliography}{99}
\bibitem{bprob}
Consider two state vectors
$\ket{a}$ and $\ket{b}$ and decompose $\ket{b}$ as
\be
\ket{b}=\alpha\ket{a}+\beta \ket{a^\perp},\nonumber
\ee
where
\be
\alpha=\braket{a}{b}\quad,\quad |\beta|=\sqrt{1-|\braket{a}{b}|^2}.\nonumber
\ee

The quantities $|\alpha|^2$ and $|\beta|^2$ have the standard probabilistic interpretation. Now,  $|\beta|$ can be viewed as the probability that $\ket{b}$ is reliably different from $\ket{a}$. To see this, recall that the optimal measurement designed to distinguish between two non-orthogonal states, such as $\ket{a}$ and $\ket{b}$, succeeds with the probability
\be
p_{succ} = \frac{1}{2} +
\frac{1}{2}\sqrt{1-|\braket{a}{b}|^2}=\frac{1}{2} + \frac{1}{2}|\beta|.\nonumber
\ee
Now, view the situation this way: with probability $p$ the two states are reliably different and with probability $1-p$ they are identical. Your probability of guessing which state you have is
\be
p_{succ}=p + (1-p)\frac{1}{2} =  \frac{1}{2} + \frac{1}{2}p.\nonumber
\ee
The comparison of the two formulae above gives $p=|\beta|$. The same argument works also for $|\alpha|$.
\end{thebibliography}

\end{document}